\documentclass[num-refs]{wiley-article}

\usepackage{graphicx}
\usepackage[space]{grffile}
\usepackage{latexsym}
\usepackage{textcomp}
\usepackage{longtable}
\usepackage{tabulary}
\usepackage{booktabs,array,multirow}
\usepackage{amsfonts,amsmath,amssymb}
\usepackage{natbib}
\usepackage{url}
\usepackage{hyperref}
\hypersetup{colorlinks=false,pdfborder={0 0 0}}
\usepackage{etoolbox}
\makeatletter
\patchcmd\@combinedblfloats{\box\@outputbox}{\unvbox\@outputbox}{
{%
  \errmessage{\noexpand\@combinedblfloats could not be patched}%
}%
\makeatother
\newif\iflatexml\latexmlfalse
}
\AtBeginDocument{\DeclareGraphicsExtensions{.pdf,.PDF,.eps,.EPS,.png,.PNG,.tif,.TIF,.jpg,.JPG,.jpeg,.JPEG}}

\usepackage[utf8]{inputenc}
\usepackage[T2A]{fontenc}
\usepackage[greek,ngerman,polish,english]{babel}

\usepackage{chemscheme}

\corremail{roberto.cerbino@univie.ac.at}

\papertype{Original Article}

\title{Differential dynamic microscopy for the characterization of polymer systems}

\author[1]{Roberto Cerbino}
\author[2]{Fabio Giavazzi}
\author[3]{Matthew E. Helgeson}

\affil[2]{University of Milan, Department of Medical Biotechnology and Translational Medicine, Via F.lli Cervi 93, 20054 Segrate, Italy}
\affil[3]{University of California Santa Barbara, Department of Chemical Engineering, 3211 Engineering II, 93106 Santa Barbara, USA}
\runningauthor{Roberto Cerbino}

\begin{document}

\maketitle
\noindent
{\textbf{1} University of Vienna, Faculty of Physics, Boltzmanngasse 5, 1090 Vienna, Austria}\\
{\textbf{2} University of Milan, Department of Medical Biotechnology and Translational Medicine, Via F.lli Cervi 93, 20054 Segrate, Italy}\\
{\textbf{3} University of California Santa Barbara, Department of Chemical Engineering, 3211 Engineering II, 93106 Santa Barbara, USA}
\selectlanguage{english}

\section*{\sffamily \large ABSTRACT}

This review summarizes recent progress in investigating polymer systems by using Differential dynamic microscopy (DDM), a rapidly emerging approach that transforms a commercial microscope by combining real-space information with the powerful capabilities of conventional light scattering analysis. DDM analysis of a single microscope movie gives access to the sample dynamics in a wide range of scattering wave-vectors, enabling contemporary polymer science experiments that would be difficult or impossible with standard light scattering techniques. Examples of application include the characterization of polymer solutions and networks, of polymer based colloidal systems, of biopolymers, and of cellular motility in polymeric fluids. Further applications of DDM to a variety of polymer systems are suggested to be just behind the corner and it is thus likely that DDM will become a tool of choice of the modern experimental polymer scientists.

\textbf{Keywords}. differential dynamic microscopy, scattering, dynamic light scattering, polymer dynamics, polymer relaxation.

\section*{\sffamily \large INTRODUCTION AND SCOPE}

The need to characterize dynamical processes and properties pervades both fundamental polymer science and applied engineering of polymeric systems. Radiation (light, x-rays and neutron) scattering has long served as a standard method in the experimental toolbox of polymer science for characterizing the dynamics of polymer molecules and materials\cite{lindner1988}. In particular, light scattering emerged as a central tool in the early era of polymer research due to its accessibility in the laboratory and versatility to different material systems. Since the early works of Zimm showing that light and other radiation scattering can be used to probe single-polymer properties and interactions,\cite{Zimm_1948} the use of scattering has expanded to probe a wide range of more complex dynamical processes including viscoelasticity,\cite{Tanaka_1973} phase instability,\cite{Hashimoto_1983} gelation\cite{Shibayama_2001} and active polymer dynamics \cite{Le_Goff_2001}.
Meanwhile, driven by parallel advances in \textit{in situ} electron \cite{Kuei_2020} and optical microscopy as well as molecular simulation over the past decades, an emerging focus of polymer research has been on the dual roles of single-molecule dynamics\cite{Moerner_2003} and structural heterogeneity\cite{Chapman_2020} on the macroscopic properties of polymers. This focus on molecular-scale processes and heterogeneity has exposed the limitations of radiation scattering and light scattering methods in particular, since although they provide access to information on the length scales of light-matter interactions, they are usually macroscopically-averaged measurements, and as such require models to extract information regarding heterogeneous processes. At the same time, despite the ability of microscopy to directly resolve microscopic processes, current real-space image analysis tools remain very limited to feature identification and single-object motion tracking, and typically require significant modification of experimental designs in order to facilitate analysis, limiting the applicability of these methods.
Fourier-domain analysis of imaging data has recently emerged as a potential way to merge the advantages of microscopy experiments and scattering measurements, while potentially avoiding some of their respective limitations \cite{Giavazzi_2014}. Specifically, converting real-space images to the Fourier domain allows one to analyze ensemble-level statistical information encoded in the entire image, while still retaining the real-space representation to assist in the modeling and interpretation of Fourier-space analysis.
In particular, differential dynamic microscopy (DDM) has emerged as an extremely powerful and versatile Fourier-domain analysis method for probing material dynamics using video microscopy data \cite{Cerbino_2008}. Although the details of the method will be discussed later, the popularized description of DDM as "dynamic light scattering on a microscope" has sparked the inspiration of researchers to apply the method to obtain dynamical information accessible to scattering measurements in the diverse contexts accessible to microscopy experiments.
This review summarizes the recent development of DDM, its principles and, importantly, its application to a growing number of contemporary areas of polymer science. Given its popular association with dynamic light scattering, it is not surprising that the development of DDM as an emerging tool for the characterization of dynamical processes in polymers exhibits a number of parallels with the development of light scattering measurements in earlier eras. However, as alluded to previously, the ability of DDM to combine Fourier-space analysis with real-space imaging information provides a number of distinct advantages which we highlight here. Based on these developments, we argue that DDM holds promise to join scattering methods as a central and multi-faceted tool in the characterization of polymeric materials and systems.

\section*{\sffamily \large BASICS OF DDM}

When using a microscope to map many small and crowded together elementary objects, one encounters two main limitations: if two or more objects are closer than the limit (roughly half of the light wavelength) imposed by diffraction, their images are indistinguishable from that of a single object; in addition, sub-wavelength objects contribute a tiny intensity signal either because they absorb/dephase little light or because they each contain a small number of primary (e.g. fluorescent) emitters. The combination of these two issues limits direct space approaches, such as particle tracking (PT) or segmentation, to being used with large and well separated objects, unless super-resolution methodologies are used \cite{Cerbino_2018}. This limitation is particularly stringent for polymeric systems, as the typical monomer size is several orders of magnitudes smaller than the wavelength of light.\selectlanguage{english}
\begin{figure}[h!]
\begin{center}
\includegraphics[width=0.70\columnwidth]{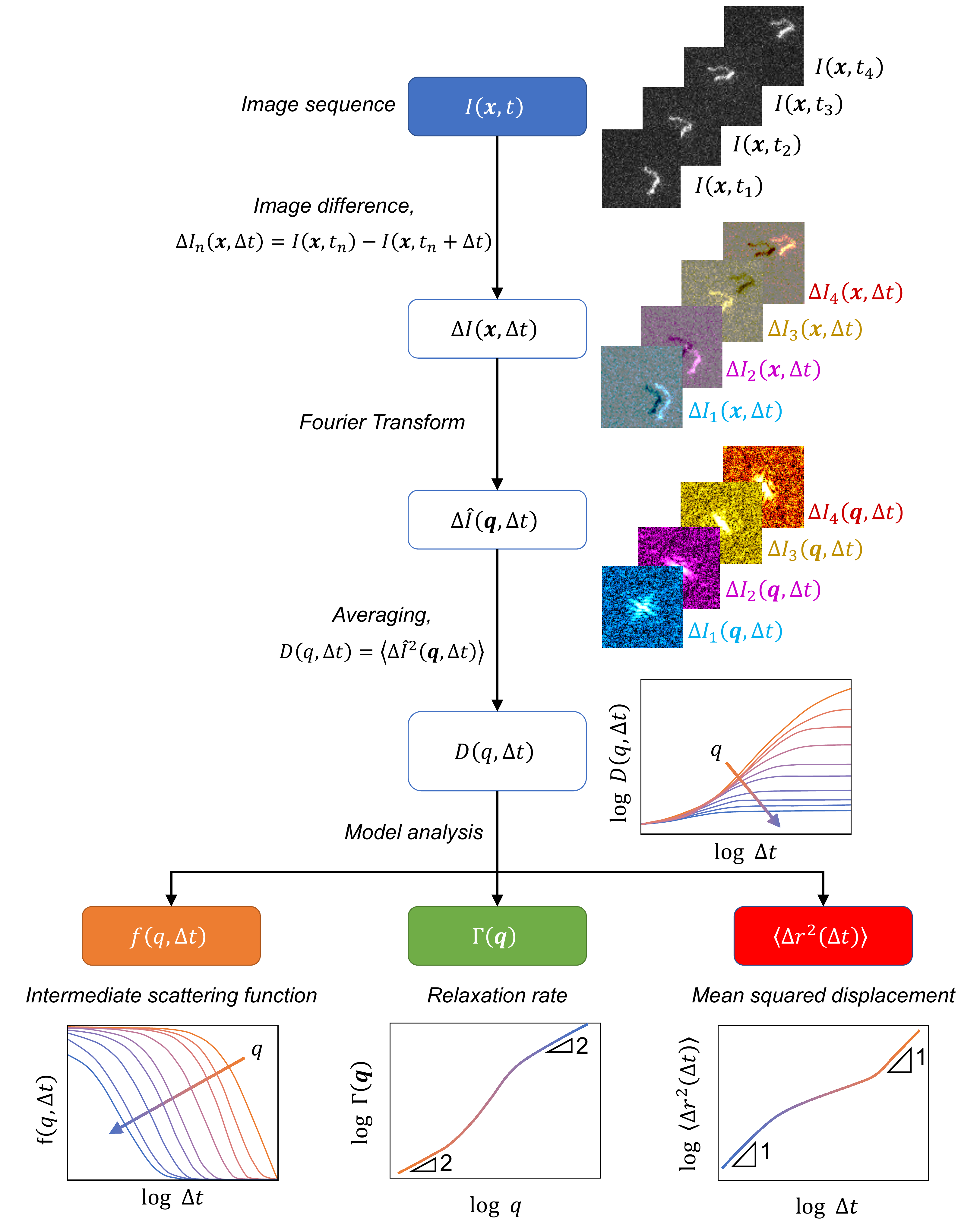}
\caption{{Typical workflow for DDM analysis of image sequences. Images and a
sketch of expected output behavior are shown for a single polymer
molecule undergoing conformational fluctuations. More details are
provided in the main text.
{\label{975224}}%
}}
\end{center}
\end{figure}

DDM provides a simple solution to this problem, as it shows that there is more to a microscopy image than meets the eye! \cite{Cerbino_2008}. We illustrate the typical working flow of DDM analysis in Fig.~\ref{975224}, which illustrates DDM for the case of a dilute, unentangled polymer solution of which we show for simplicity only one single chain, whose microscope image $I(\mathbf{x},t)$ is collected at various times ($t_{1},t_{2},t_{3},t_{4}$ in the figure). 

If one considers a generic image $I(\mathbf{x},t)=I_{BG}(\mathbf{x})+\delta i(\mathbf{x},t)$ acquired at time t, the tiny signal $\delta i(x,t)$ due to the temporally and spatially fluctuating distribution of the small elementary objects can be isolated from the time-independent background $I_{BG}(\mathbf{x})$ by calculating the image difference $\Delta I_{n}(\mathbf{x},\Delta t)=i(\mathbf{x},t_{n})-i(\mathbf{x},t_{n}+\Delta t)=\delta i(\mathbf{x},t_{n})-\delta i(\mathbf{x},t_{n}+\Delta t)$. The variable $\Delta I_{n}(\mathbf{x},\Delta t)$ is a stochastic process that fluctuates both in time and space. However, due to the limited spatial resolution of the optical microscope and to the two-dimensional (2D) nature of the image, the relationship between the intensity at each point $\mathbf{x}=(x,y)$ of the detector and the concentration $c(\mathbf{x},z;t)$ of the elementary objects that make up the three-dimensional (3D) sample does not allow for a simple analysis of the signal in direct space. DDM achieves the trick by performing the calculation of the ensemble image temporal correlation functions in the reciprocal space.

To this aim, each image difference $\Delta I_{n}(\mathbf{x},\Delta t)$ is first Fourier transformed to obtain $\Delta \hat{I}_{n}(\mathbf{q},\Delta t)=FT_{2D}[\Delta I_{n}(\mathbf{x},\Delta t)]$, where $FT_{2D}[...]$ is the the 2D spatial Fourier transform ($\mathbf{x}\to \mathbf{q}$). Under very reasonable experimental hypotheses \cite{Giavazzi2009}, the image structure function $D(\mathbf{q},\Delta t)=<|\Delta\hat{I}_{n}(\mathbf{q},\Delta t)|^{2}>_{t_{n}}$ can be calculated, where the average $\left<...\right >_{t_{n}}$ is performed over realizations with the same $\Delta t$ but different $t_{n}$. One has that \cite{Giavazzi_2014}:

\begin{equation}
\label{eq:ddmeq}
D(\mathbf{q},\Delta t)=A(\mathbf{q})[1-f_{\mathbb{R}}(\mathbf{q},\Delta t)] + B(\mathbf{q})
\end{equation}

where $f_{\mathbb{R}}(\mathbf{q},\Delta t)$ is the real part of the normalized intermediate scattering function $f(\mathbf{q},\Delta t)$, the term $B(q)$ accounts for the detection noise, and the term $A(q)$ is a static amplitude term that depends on the contrast mechanism behind the image formation and on the distribution and shape of the elementary objects \cite{Giavazzi2009,Giavazzi_2014}.

The normalized intermediate scattering function is ordinarily probed by Dynamic Light Scattering (DLS), a technique that is routinely employed for the characterization of polymers in solutions \cite{Burchard}. Eq. \ref{eq:ddmeq} implies that any existing model for the intermediate scattering function, developed for instance for DLS applications, can be used directly to fit DDM data. Several models exist to describe the intermediate scattering function of dilute, semi-dilute and concentrated polymer solutions. The interested reader can find a good entry point to the topic in \cite{colby2003} and a more advanced treatment in \cite{Han_2011}. Here we briefly illustrate the simplest case \textit{i.e.} that of a dilute, unentangled polymer solution, which is pictorially represented in Fig.\ref{975224}. For a polymer with statistical segment length $b$ and gyration radius $R_{g}\gg b$, the dynamics exhibits three different regimes as a function of $q$: for $q\ll 1/R_{g}$ one has $f(q,\Delta t)=e^{-\Gamma(q)\Delta t}$, whose relaxation rate $\Gamma(q)=D_{t}q^{2}$ captures the Brownian motion with diffusion coefficient $D_{t}$ of the center of mass of the chain; for $q\gg 1/b$, a diffusive dispersion relation $\Gamma(q)=D_{m}q^{2}$ is predicted, where $D_{m}\gg D_{t}$ is associated to the diffusion of the monomers; for intermediate $q$ ($1/R_{g}<q<1/b$), the internal chain dynamics is probed and one expects a dispersion relation $\Gamma\sim q^4$ in the absence of hydrodynamic interactions (Rouse limit, see Fig.\ref{975224}) or, if hydrodynamic interactions are considered, the weaker scaling $\Gamma\sim q^3$ (Rouse-Zimm limit). Interestingly, whenever a model is available to connect the intermediate scattering function with the mean-squared-displacement of the scatterers, it is also possible to extract the latter from DDM experiments, and one would observe again three different regimes: diffusive dynamics for short and long times, and sub-diffusion for intermediate ones. It must be stressed that, access to all these regimes of $q$ and $t$ with DDM depends on how the characteristic length and time scales of the polymer system compare with the experimentally accessible wave-vector range, camera acquisition frame rate and overall experimental duration.

Eq. \ref{eq:ddmeq} also implies that a microscope can be used as a quantitative DLS instrument that probes simultaneously many wavevectors in the range $[q_{min},q_{max}]$ where $q_{min}$ is set by the size of the imaged sample portion and $q_{max}$ is set by the pixel size, in both cases after suitable magnification. Under typical circumstances, one can cover the range $[0.1,10]$ $\mu$m$^{-1}$, where some improvement may be obtained by performing experiments with different magnifications. This wavevector range extends on the lower side what is accessible with commercial DLS instruments, which makes DDM a useful complement to DLS. On the temporal side, it must be stressed that the temporal averaging leading to Eq. \ref{eq:ddmeq} requires that the sample dynamics can be considered stationary (or quasi-stationary) during the time-window over which the DDM analysis is performed. When this is not the case, it is common practice to limit the duration of an image sequence so that the quasi-stationarity hypothesis holds and acquire several of such image sequences, each one being representative of a given "age" of the system under study. Each sequence is then analyzed with DDM to characterize the sample dynamics at a given age. While this strategy has been proven to work in several cases \cite{Ferri_2011,Cho_2020,Giavazzi_2020}, it poses some obvious limitations on the capability of DDM to follow non-stationary dynamics, a problem that notoriously affects also DLS. 

A notable advantage of DDM over DLS is that the former can operate with a variety of microscopy contrast mechanisms including bright-field, phase-contrast, dark-field, differential interference contrast, and depolarized scattering-based microscopy or wide-field, confocal, light-sheet fluorescent based microscopy (see \cite{Giavazzi_2014} and \cite{Cerbino2017} for a comprehensive review). The capability of using fluorescence as a contrast mechanism makes DDM akin to Fluctuation Correlation Spectroscopy (FCS), a technique that is increasingly being applied to the study of polymer systems \cite{W_ll_2014}. It must be stressed, however, that FCS does not provide wave-vector resolved information. The reader interested in a more detailed discussion on the relationship between fluorescence-based DDM and other FCS-like techniques can refer to Ref. \cite{Giavazzi_2014}.
 
Two very recent studies address the potential of DDM for diffusion characterization \cite{Struntz_2018} and particle sizing \cite{Eitel_2020} compared to DLS, FCS, PT, TEM, SEM, turbidity measurement and static light scattering (SLS). DDM, the youngest among the mentioned approaches, can provide very reliable results and offers the unique possibility to operate in highly turbid samples, an advantage that comes either from the partial coherence of the typical microscope light sources \cite{Giavazzi2009} or from the optical sectioning capability of confocal \cite{Lu_2012} and light-sheet \cite{Wulstein_2016,Struntz_2018} microscopes. 

Before turning to the description of applications of DDM, we would like to stress that, while DDM analysis provides as an immediate output the structure function $D(q,\Delta t)$, extraction of other quantities (Fig.\ref{975224}), such as the intermediate scattering function $f(q,\Delta t)$, the relaxation rate $\Gamma(q)$ (here we assume for simplicity that there is only one dominant relaxation process), and the mean squared displacement $\langle\Delta r^2(\Delta t)\rangle$ is less immediate and requires often additional information and/or assumptions. In brief, if a model for $f(q,\Delta t)$ is available, $A(q)$, $B(q)$, and $\Gamma(q)$ can be simply obtained with a fitting procedure \cite{Cerbino_2008,Giavazzi2009}. If this is not the case, knowledge of $A(q)$ and $B(q)$ is needed to extract $f(q,\Delta t)$. This typically requires a large acquisition frame rate, to efficiently decouple the detection noise from the genuine sample signal, and a sufficiently long acquisition, to capture full sample decorrelation. In all the cases in which the main contribution to $B(q)$ is shot-noise, the latter can be determined in principle by prior calibration of the detector without sample and under similar illumination conditions. Alternatively and more simply, one can rely on the fact that $B(q)$ is q-independent and $A(q)\rightarrow 0$ for large $q$. If the decay of $A(q)$ to zero occurs for $q<q_{max}$ (\textit{i.e.} if the pixel size is smaller than the optical resolution), one can determine $B(q)$ as the high-q limit of $D(q,\Delta t)$\cite{Giavazzi_2020}. In Ref.\cite{Giavazzi_2020}, it is also shown how $A(q)$ can be obtained from the time averaged Fourier power spectrum of individual images, which requires that the intensity background $I_{BG}(\mathbf{x})$ is spatially homogeneous. Extracting the mean squared displacement $\langle\Delta r^2(\Delta t)\rangle$ from $f(q,\Delta t)$ requires the validity of additional hypotheses about the statistical properties of the particles motion \cite{Edera_2017,Bayles_2017,Escobedo_S_nchez_2018}.

The above discussion shows that, when presenting the results of DDM analysis, an effort should be made to make available to the reader all the relevant intermediate steps of the analysis: for example, showing results for the relaxation rate $\Gamma(q)$ without showing typical structure functions $D(q,\Delta t)$ and the corresponding  fitting curves (at least for some values of $q$) should be avoided.

\section*{APPLICATIONS}

In this Section, we describe four key areas in which DDM has already unveiled its potential to deal with challenging samples and experimental configurations that involve polymers.

\subsection*{Characterization of polymer solutions and networks}\selectlanguage{english}
\begin{figure}[h!]
\begin{center}
\includegraphics[width=0.70\columnwidth]{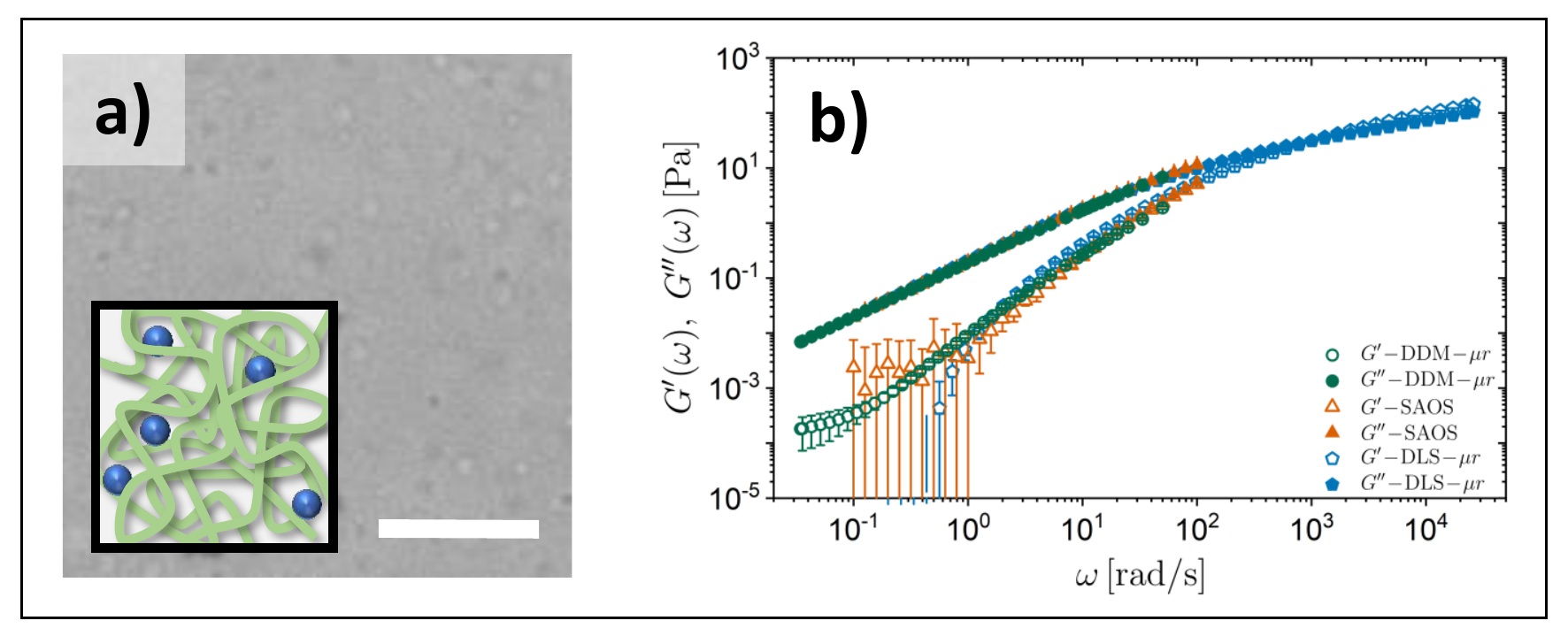}
\caption{{DDM-microrheology of complex fluids~\protect\cite{Escobedo_S_nchez_2018} . (a)
Representative bright-field image of a solution of PEO in water with
embedded polystyrene spheres (diameter 330 nm).~The scale bar indicates
10 \selectlanguage{greek}μ\selectlanguage{english}m.~ In the inset, a pictorial representation of the tracer particles
(in blue) embedded in the polymeric solution is shown. In panel (b), the
frequency-dependent viscoelastic moduli G'(\selectlanguage{greek}ω\selectlanguage{english}) and G''(\selectlanguage{greek}ω\selectlanguage{english}) of an aqueous
PEO solution measured with DDM microrheology (green red circles) are
compared with the ones determined~\emph{via} conventional
small-amplitude oscillatory shear (SAOS) rheology (orange upright
triangles) and DLS (blue pentagons) measurements, respectively.
DLS-based microrheology was performed with the same tracer particles
used in the DDM experiment. Image courtesy:~M. A. Escobedo-Sánchez.
{\label{309344}}%
}\selectlanguage{ngerman}}
\end{center}
\end{figure}\selectlanguage{ngerman}

Many early applications of DDM mirrored the development of DLS for probing dynamics in polymeric systems, including the measurement of diffusion coefficients and their dependent properties. For example, a recent study employed a combination of fluorescence correlation spectroscopy (FCS) and DDM to characterize long-time diffusion in concentrated polymer solutions \cite{Shokeen_2017}, and the results confirmed predictions from theoretical scaling relations for entangled polymers.
More recently, it was demonstrated that DDM can be used to perform passive probe microrheology experiments, in which a material of interest is seeded with spherical Brownian particles, and their motion is used to extract linear viscoelastic properties of the matrix material. Employing theories conventionally used for analysis of DLS microrheology \cite{Furst_2018}, one can extract the mean-squared displacement (MSD) of embedded probes, and subsequently the frequency-dependent viscoelastic storage (G\selectlanguage{english}') and loss (G\selectlanguage{english}") moduli via the Generalized Stokes-Einstein relation \cite{Bayles_2017,Edera_2017}. Initial demonstrations on semi-dilute polymer solutions \cite{Edera_2017,Escobedo_S_nchez_2018} and micellar fluids \cite{Bayles_2017} showed that DDM microrheology can quantitatively reproduce the MSD and linear viscoelasticity obtained by both conventional multiple particle tracking (MPT) microrheology as well as bulk rheological characterization (see Fig. \ref{309344}). Significantly, it was shown that DDM microrheology could be employed under conditions where MPT is inaccessible due to imaging limitations \cite{Bayles_2017}.
This superior versatility of DDM has been exploited to characterize and screen complex material formulations including polymer networks that would otherwise be intractable using conventional microrheology methods. For example, using Brownian probe measurements on a series of crosslinking polyacrylamide solutions, DDM microrheology was used to verify the concept of time-cure superposition \cite{Winter_1986,Larsen_2008} and its use in precise estimation of critical gelation exponents and the gel point \cite{Bayles_2017}. More recently, such experiments were implemented in a novel combination of automated sample preparation, microscopy, DDM analysis and machine learning to perform high-throughput screening of gelation kinetics of silk fibroin biopolymer networks over a wide, multi-component compositional space in order to identify compositional windows with desirable gelation times \cite{learning}. Such integrated measurements and methods involving DDM microrheology show significant promise for the future application of DDM microrheology to polymeric materials with complex composition and design spaces.

\subsection*{Characterization of polymer based colloidal systems}\selectlanguage{english}
\begin{figure}[h!]
\begin{center}
\includegraphics[width=0.70\columnwidth]{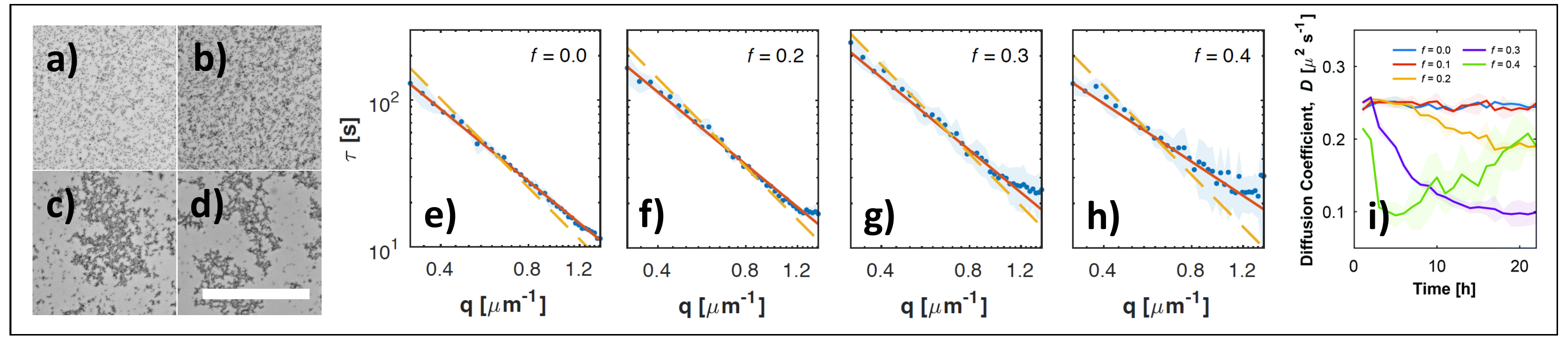}
\caption{{Monitoring particle aggregation with DDM. (a-d) Representative
bright-field images of DNA-coated silica microparticles with a variable
fraction of linkers (f = 0, 0.2, 0.3, 0.4, respectively) at the end of
an aggregation experiment. (e-h) Relaxation time \selectlanguage{greek}τ \selectlanguage{english}as a function of the
wave vector q, as obtained from DDM analysis performed on the same
samples shown in panel (a-d) (Blue points). The solid orange line
indicates the best power law fit~\(\tau\propto q^{-\alpha}\), while the dashed
yellow line corresponds to the best Brownian fit \(\selectlanguage{greek}τ \selectlanguage{english}= Dq^{-2}\). The
latter is used to extract the effective diffusion coefficient D at
different time points, as shown in panel (i). For f\textless{}0.2, the
effective diffusion coefficient remains roughly constant over time,
indicating the stability of the ``colloidal gas'' phase. The rapid
decrease in D~observed for larger fractions of linkers, corresponds to
the formation of large aggregates.~ Adapted from~\protect\cite{Lanfranco_2020}
under Creative Commons Attribution-NonCommercial 3.0 Unported Licence. ~
{\label{663758}}%
}}
\end{center}
\end{figure}

Synthetic polymer latex particles prepared via emulsion polymerization are routinely characterized with DDM, typical examples being polystyrene (PS) \cite{Cerbino_2008} and Poly(methyl methacrylate) (PMMA) \cite{Lu_2012}. In addition to the particle translational diffusivity, DDM is sensitive to a variety of dynamics of polymeric particles, including their rotational Brownian motion \cite{Giavazzi_2016} and directed motion \cite{Dienerowitz_2013}.
Beyond this simple role as building blocks of colloidal particles, polymers are frequently used in combination with colloids, mostly as depletion \cite{Lekkerkerker_2011} or surface-functionalization agents \cite{2013}. Lanfranco \textit{et al.} \cite{Lanfranco_2020} used DDM to measure the diffusivity of DNA-coated colloidal particles and monitor in time their aggregation state induced by selective DNA-DNA interactions (see Fig. \ref{663758}).
In addition to equilibrium studies DDM has been used to characterise non-equilibrium processes in polymer-colloids. Gao \textit{et al.} \cite{Gao_2015} used DDM in combination with other techniques to study simultaneous spinodal decomposition and physical gelation of a colloidal system of nanoemulsion droplets in the presence of thermoresponsive polymers. Access to very small q-values highlighted a combination of fast diffusive dynamics within the colloid-rich domains, and slow, intermittent directional motion of individual domains driven by spinodal coarsening. Wang \textit{et al.} \cite{Wang_2019} employed DDM to study capillary waves arising in a colloid-polymer mixture that exhibits liquid-gas phase separation. They obtained an accurate estimate of the capillary velocity for several samples in which the concentration of the particles and the polymer was varied, finding values consistent with prior studies. Cho \textit{et al.} \cite{Cho_2020} investigated aggregation, geometric percolation, and the subsequent transition to nonergodic dynamics in a system of polystyrene-poly(N-isopropylacrylamide) (PS-PNIPAM) core-shell colloidal particles synthesized by emulsion polymerization. By combining different image acquisitions on samples with different particle concentration, they highlighted the existence of three distinct regimes in the formed gels: for small q, the dynamics is overdamped and similar to that of a homogeneous viscoelastic medium; for  intermediate q, a q-independent dynamics dominated by the density fluctuations at the length scale of the clusters; for large q, the internal vibrations of the fractal clusters are measured. 
Finally, Sentjabrskaja \textit{et al.} \cite{Sentjabrskaja_2016} employed DDM with a system made of two species of colloidal particles to study the collective dynamics of the smaller ones (intruders) in the mobile, crowded environment (matrix) landscaped by the larger ones. Use of two-color fluorescent tagging of the particles, allowed the authors to investigate separately the dynamics of the two species, providing information that would be simply inaccessible with DLS. The experiments, conducted for different ratios of the particles size, revealed extended anomalous dynamics for specific values of the size asymmetry and of the probed length scale. In particular, a logarithmic decay of the collective intermediate scattering function of the intruders was observed at length scales comparable to the size of the matrix particles. An outcome of this study was the identification of a critical size ratio $\delta c$ below which the intruders diffusively move in a glassy porous matrix, and above which crowding leads to a glassy dynamics of the intruders themselves. 

\subsection*{Characterization of biopolymers}\selectlanguage{english}
\begin{figure}[h!]
\begin{center}
\includegraphics[width=0.70\columnwidth]{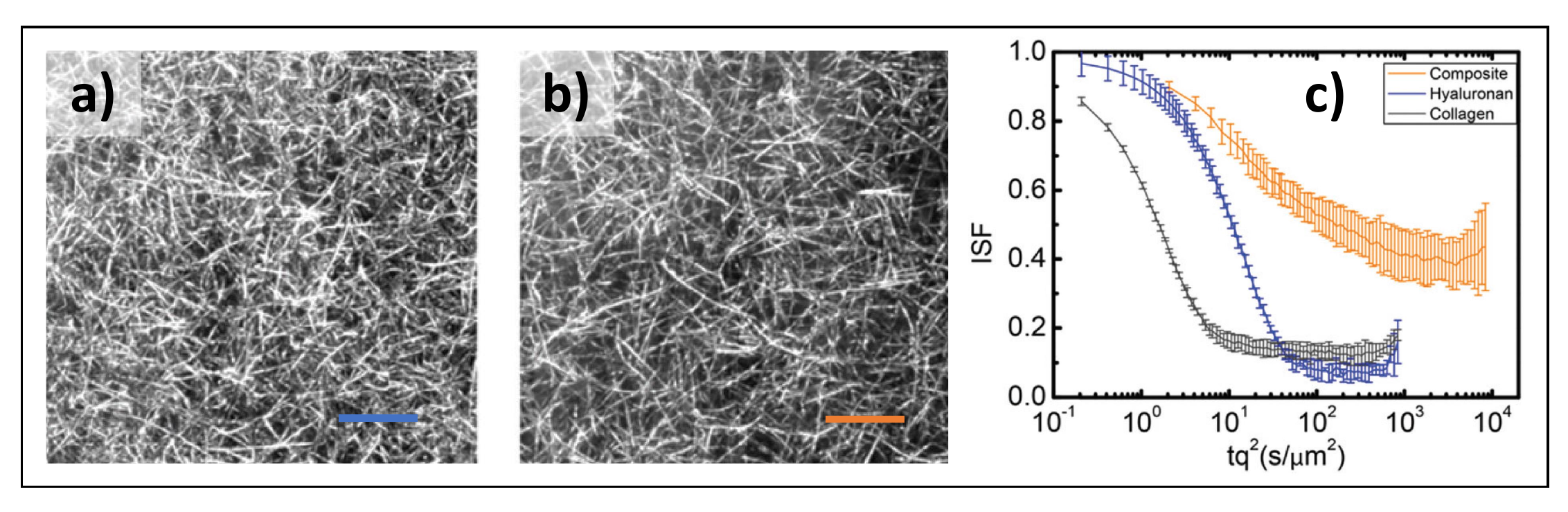}
\caption{{Tracking-free determination of particle dynamics in composite
collagen--hyaluronan networks.~ (b and c) Confocal images of the
fibrillar collagen network for a 1 mg mL\textsuperscript{-1}~pure
collagen network (b) and for a composite where the collagen fibrils are
embedded in a hyaluronan background network that is not visible in the
image (c). Scale bars indicate 10 \selectlanguage{greek}μ\selectlanguage{english}m.~ (c) Intermediate scattering
functions (ISFs) obtained from DDM analysis of bright-field image
sequences of different samples where 0.6 \selectlanguage{greek}μ\selectlanguage{english}m particles are dispersed: a 1
mg mL\textsuperscript{-1}~collagen (black curve) network, a 2 mg
mL\textsuperscript{-1}~hyaluronan sample (blue), and composite
collagen--hyaluronan network (orange). Adapted from~ ~\protect\cite{Burla_2020}
under Creative Commons Attribution-NonCommercial 3.0 Unported Licence.
{\label{777003}}%
}}
\end{center}
\end{figure}

Biological polymers and systems have become one of the most rapidly expanding areas of application for DDM. The present review will focus primarily on biopolymer materials reconstituted outside their native environment; there are also a growing number of examples involving studies on living cells and tissues, which will be described in the following section.
At the molecular level, researchers have employed DDM to characterize the rheological properties and condensation of soluble proteins in solution in a manner analogous to more conventional DLS measurements. Model studies on solutions of lysozyme and hemoglobin A \cite{Safari_2015} demonstrated that the measured DDM signal of protein condensates could be accurately described using the cumulant expansion commonly employed in DLS, resulting in estimation of size distributions of condensate droplets.  Alternatively, DDM microrheology has been employed to understand the role of protein composition on the viscosity of whole mouth saliva samples \cite{Pushpass_2019}, as well as the mobility of ribonucleoprotein granules within Drosophilia oocytes \cite{Fajner_2020}. The latter studies highlight the utility of DDM for characterizing rheological information of scarce materials and \textit{in vivo} systems that would be difficult if not impossible to obtain using conventional methods.
DDM has also been used as an effective probe to study the formation and properties of biopolymer networks. Recently, DDM microrheology was used as an \textit{in situ} probe of the pH-dependent gelation of hyaluronan networks, and was used to show that the emergence of elasticity was concomitant with expulsion of water from the newly formed network, as confirmed using independent studies of water dynamics \cite{van_Dam_2020}. Similar DDM microrheology studies on the formation of hyaluronic acid networks tracked the evolution of dynamics during both physical and chemical gelation \cite{Burla_2020}. Interestingly, whereas chemically-crosslinked hyaluronic acid networks exhibited sub-diffusive and non-ergodic probe motion consistent with an elastic medium, physically-crosslinked networks exhibited heterogeneous dynamics including a population of purely viscous motion that is unobservable in bulk rheological measurements (see Fig. \ref{777003}). These studies demonstrate the importance of quantifying micro-scale rheology and dynamics to probe material heterogeneities and the utility of DDM in probing them.  
Building toward more complex and native structured biomaterials, a number of recent studies have employed DDM to isolate the role of various components and processes in the dynamics and mechanics of cytoskeletal networks involving actin and microtubule assemblies. For example, recent multi-channel fluorescence DDM experiments \cite{Lee_2021} identified the role of microtubules in moderating ballistic motion and contractile forces associated with myosin-driven activity of reconstituted cytoskeletal networks. It was observed that this activity-mediated forcing of the network leads to a transition in the anomalous dynamics of particles embedded in the network from sub-diffusive motion at short times due to the interplay of diffusion with thermal fluctuations of the network \cite{Anderson_2019,Lee_2021}, toward ballistic motion at long times due to the influence of myosin-driven active motion of the network \cite{Lee_2021}.
Other studies have examined the influence of these complex dynamics of actin-microtubule networks on various biological transport processes. For example, DDM has been combined with single molecule tracking showed that cytoskeletal crowding and sub-diffusive transport lead to increased compaction and conformational rigidity of DNA in reconstituted actin-microtubule networks \cite{Regan_2019,Wulstein_2019}. Another study showed that in vivo cytoskeletal vesicle transport in Drosophila oocytes could be directly linked to a combination of thermal fluctuations and advective motion associated with active actin motion \cite{Drechsler_2017}. An important feature common to this growing body of studies on cytoskeletal networks is that because DDM relies on statistical correlations of fluctuations across an entire imaging plane, dynamical information can be extracted without tracking individual molecules, circumventing the need for super-resolution optical methods and enabling studies in complex biological milieu.

\subsection*{Cellular motility in polymer solutions}\selectlanguage{english}
\begin{figure}[h!]
\begin{center}
\includegraphics[width=0.70\columnwidth]{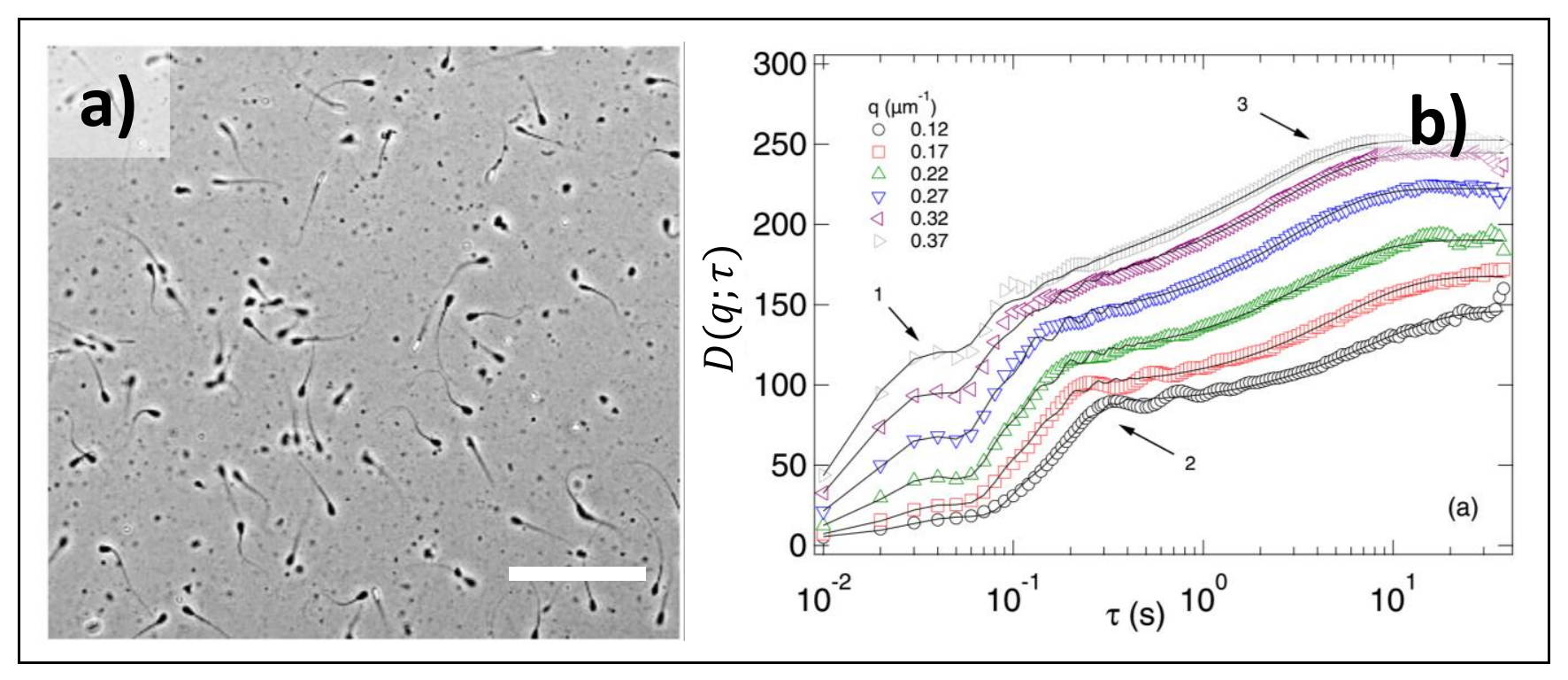}
\caption{{High-throughput characterization of the motility of spermatozoa. (a)
Representative phase-contrast image of a diluted thawed bull semen
sample with ~ 20 × 10\textsuperscript{6} cell/ml. Scale bar = 100 \selectlanguage{greek}μ\selectlanguage{english}m.(b)
Image structure function obtained from DDM analysis~at 6 values of q,
specified in the legend. Black lines are best fitting curves with a
model incorporating the main contributions to the sperm cells dynamics,
namely, head oscillation (1), swimming (2), and diffusion of non-motile
cells (3). Adapted from~\protect\cite{Jepson_2019}~ under Creative Commons
Attribution Licence.
{\label{784100}}%
}\selectlanguage{ngerman}}
\end{center}
\end{figure}

The potential of DDM in providing a statistically robust estimate of the parameters describing cell motility at different scales has been demonstrated on a variety of systems, spanning  from diluted suspensions of flagellated bacteria or algae \cite{Martinez_2012} to confluent monolayers of epithelial cells \cite{Giavazzi_2018,Chioccioli_2017}. DDM works even in thick 3D samples and for relatively high cell concentrations, a regime which is not easily accessible with standard optical methods, based for example on particle tracking or DLS.
In many biologically relevant situations, cell motility occurs in complex, heterogeneous environments, often displaying non-trivial, length scale-dependent rheology, making the interaction between motile cells and polymer solutions a topic of great applicative relevance.
A very nice example along this line is provided by Ref. \cite{Martinez_2014}. There, Martinez \textit{et al.} investigate the swimming behaviour of E. coli in solutions of both linear (polyvinylpyrrolidone (PVP)) and branched (Ficoll) polymers over a wide range of  molecular weights and concentrations. In this work, by combining DDM with dark-field flicker microscopy (DFFM) the Authors find that, for almost all considered samples, the swimming behavior is fully compatible with the motion in a Newtonian fluid characterized by an effective viscosity significantly smaller than the one of the polymer solution. This result can be accounted for by a simple model where the flagella, by moving in a short polymer-depleted nanochannels "dug" by their own rapid motion, experiences a viscosity close to the one of the pure buffer. Besides their intrinsic interest, these exciting results demonstrate the potential of DDM of exploiting motile  bacteria as "active nanorheometers" to probe the rheology of complex fluids at the sub-micrometer scale.
A completely different regime of cell-polymer interaction is the one where the motile cells are suspended in a diluted solution of polymers much smaller than the cell size. In this case, the polymer primarily acts as a depletion agent, promoting the aggregation of the cells. This problem is theoretically and experimentally explored in Ref. \cite{Schwarz_Linek_2012}, where  a dilute suspension of motile E. coli is mixed with sodium polystyrene sulfonate (NaPSS, molecular weight 64700 g/mol) at different concentrations. Polymer-induced aggregation of bacteria is shown to promote the formation of  clusters showing persistent collective rotation due to the non-vanishing total torque exerted by the bacteria at its boundaries. In this work, DDM was used to monitor the motility of freely swimming bacteria.
More recently, DDM has been used for characterizing the motility of spermatozoa in both fresh and defrosted samples of bull semen at different dilutions \cite{Jepson_2019}. Exploiting the intrinsic multiscale capability of DDM, and in particular its ability to access small wavevectors, Jepson \textit{et al.} were able to separate and measure the many different contributions to the dynamics exhibited by these samples, where motile and non-motile sperm cell are dispersed in a complex fluid matrix (the seminal plasma) (see Fig. \ref{784100}). The obtained motility parameters are in good agreement with the ones obtained with single particle tracking, which is the  current method of choice for evaluating spermatozoon motility in a lab setting. These promising results suggest tha DDM could be successfully used in different steps of the process of fertility assessment in veterinary practice, both on-farm and in-lab.
A paradigmatic example of how the chemical composition and the rheological properties of a polymeric solution are key in determining the spatio-temporal behaviour of motile cellular structures is represented by the interaction of ciliated cells, like the ones lining the airway epithelium, with mucus. In Ref. \cite{Chioccioli_2019} it is shown that, beside the ciliary beat frequency (CBT) also the extent of spatial correlations in the collective ciliary beating dynamics is a sensitive and robust readout of the ability of cells to promote mucociliary clearance. In this work, the spatio-temporal features of the collective ciliary dynamics exhibited by  in-vitro samples of live human bronchial epithelial ciliated cells are captured via an automated image analysis procedure based on a sequence of DDM analyses performed over subregions of different size. Multi-scale DDM (multi-DMM \cite{Feriani_2017}) enables the determination of both CBT and of the spatial correlation length scale over which the beating dynamics is coherent, providing accurate phenotyping of cultured cells and enabling a quantitative assessment of the efficacy of pharmaceutical treatments.

\section*{CONCLUSIONS AND PERSPECTIVES}
The present review has provided an overview of recent progress in the investigation of polymeric systems enabled by the use of DDM.
Besides providing a convenient and accessible alternative to traditional light scattering methods, DDM opens up a number of exciting and largely unexplored possibilities, leveraging its unique combination of features.
For example, its intrinsic user-independence, the compatibility with different imaging modalities and the robustness against optical imperfections and multiple scattering make DDM an ideal candidate for the integration in automated platforms, combining sample preparation, microscopy, quantitative image analysis and machine learning, performing high-throughput characterization of polymeric materials with complex composition and design spaces \cite{learning}.
In general, the fact that DDM is compatible with a variety of imaging modes, can be exploited to simplify the design of experiments aimed at simultaneously monitoring different components within a given system. For example, multimodal imaging and DDM can be combined to study the dynamical interplay of different substructures within a composite material \cite{Drechsler_2017} or to perform a tracer-based microrheology experiment while simultaneously measuring the spontaneous relaxation dynamics of the matrix.
The combination of DDM with a shear cell or a flow cell enabling optical access to the sample \cite{Aime_2019} or its integration in a rheo-microscopy setup \cite{Burroughs_2020} would enable the time- and space-resolved observation of samples under the application of controlled stresses and deformations, providing an insight in the microscopic events underlying the macroscopic mechanical response of a material. In this kind of application, it is particularly important to be able to resolve localized events occurring in the material and account for the presence of subregions showing distinct behaviour and microstructure. DDM, as an imaging-based technique, can be easily adapted to the study of such spatially heterogeneous samples. This can be obtained for example by dividing a large field of view into smaller regions of interest (ROIs) that are analyzed separately, or by combining observations performed with different objective magnifications \cite{Feriani_2017}, to optimize trade-offs between spatial resolution, spectral resolution, and statistical robustness.
The use of DDM as a diagnostic tool for quasi-real-time monitoring of samples during production, processing, transport or aging is fostered by the availability of optimized software implementations, enabling the quasi-real time DDM analysis of an acquired image sequence \cite{time}. Moreover, the continuous progress in imaging sensors technology and the introduction, in combination with DDM, of clever acquisition/illumination schemes will enable access to unprecedentedly fast dynamics \cite{Arko_2019,You_2021}.
DDM is robust against multiple scattering compared to DLS, and has been exploited by many investigators to study samples that would otherwise have been impossible with optical methods. However, a systematic study of the effects of multiple scattering on DDM analysis is still missing, and understanding the associated limitations could be important for extending DDM to a wider class of samples and experiments. Finally, we note that given the contemporary development of new data exploration methods for molecular simulation, there may ultimately be unexplored utility in applying DDM to visualized dynamic simulations, which may ultimately allow for accelerated computation of dynamical properties that may otherwise be expensive to access, such as the intermediate scattering function. Overall, we believe these recent and potential new developments for DDM provide a powerful new tool for exploring emerging frontiers in polymer science and engineering.

\subsection*{ACKNOWLEDGMENTS}

We kindly thank Charles Schroeder and Peter Zhou for providing microscopy images for Figure \ref{975224}. We also thank M.A. Escobedo-S\selectlanguage{ngerman}ánchez for his help in the preparation of Fig. \ref{309344}. MEH acknowledges partial support from the BioPACIFIC Materials Innovation Platform of the National Science Foundation under Award No. DMR-1933487. FG acknowledges funding from from the Associazione Italiana per la Ricerca sul Cancro (AIRC) - Project MFAG 22083.
\clearpage

\subsection*{AUTHOR BIOGRAPHIES}\selectlanguage{english}
\begin{figure}[h!]
\begin{center}
\includegraphics[width=0.14\columnwidth]{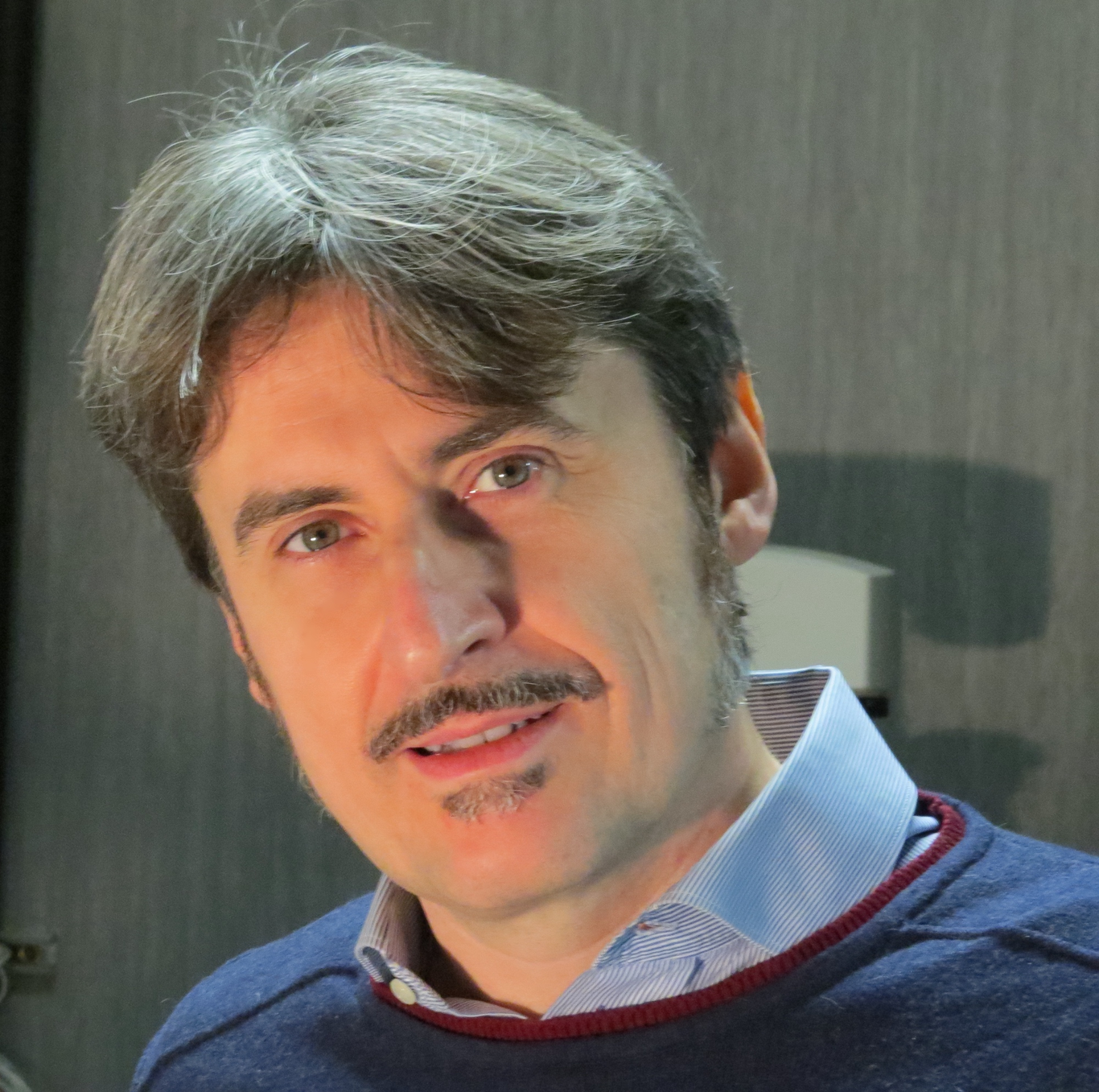}
\caption{\selectlanguage{polish}{\textbf{Roberto Cerbino}~was born in 1975 in Italy. In 2001, he got a
Laurea in Physics~\emph{summa cum laude} from the University of Milano
(Italy), where he also obtained his PhD in Applied Physics in 2004.
After a Marie Skłodowska-Curie fellowship at the University of Fribourg
(Switzerland), he became tenured at the University of Milano (Italy) in
2007. Visiting researcher at the University of Ottawa (Canada) in 2010
and visiting professor at the École normale supérieure de Lyon (France)
in 2018, since March 2021, he is professor of Experimental Soft Matter
Physics at the University of Vienna (Austria).~ His domain of research
and expertise is the physics of soft and biological materials,
investigated with innovative opto-mechanical tools.
{\label{533251}}%
}\selectlanguage{ngerman}}
\end{center}
\end{figure}\selectlanguage{ngerman}\selectlanguage{english}
\begin{figure}[h!]
\begin{center}
\includegraphics[width=0.14\columnwidth]{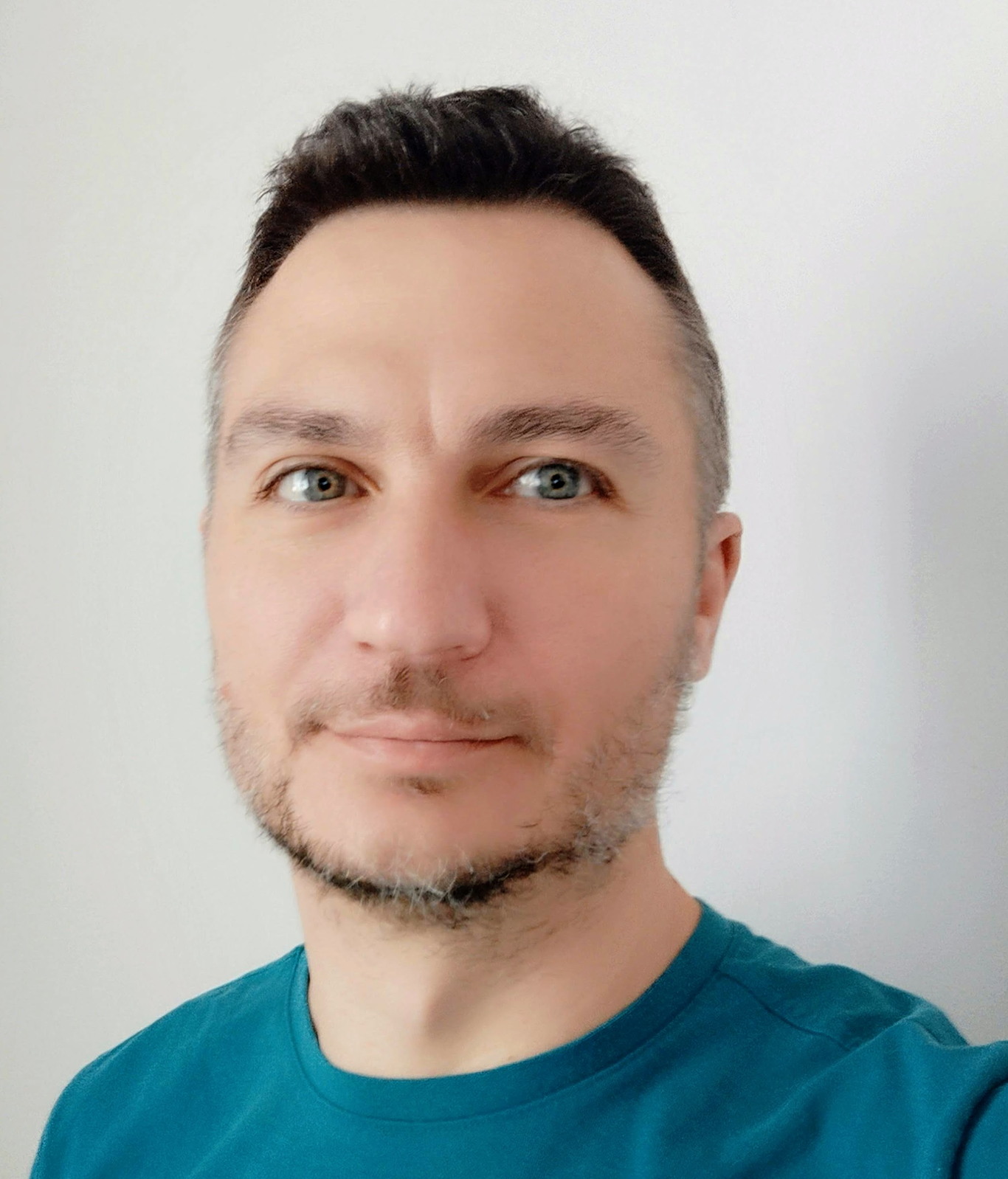}
\caption{{\textbf{Fabio Giavazzi~}holds a Ph.D. in Physics and is currently a
research fellow and tenure-track assistant professor of Applied Physics
at the University of Milan (Italy). His research activity is focused on
the development and the application of quantitative optical microscopy
methods for the investigation of bio-soft matter systems.
{\label{224815}}%
}}
\end{center}
\end{figure}\selectlanguage{english}
\begin{figure}[h!]
\begin{center}
\includegraphics[width=0.14\columnwidth]{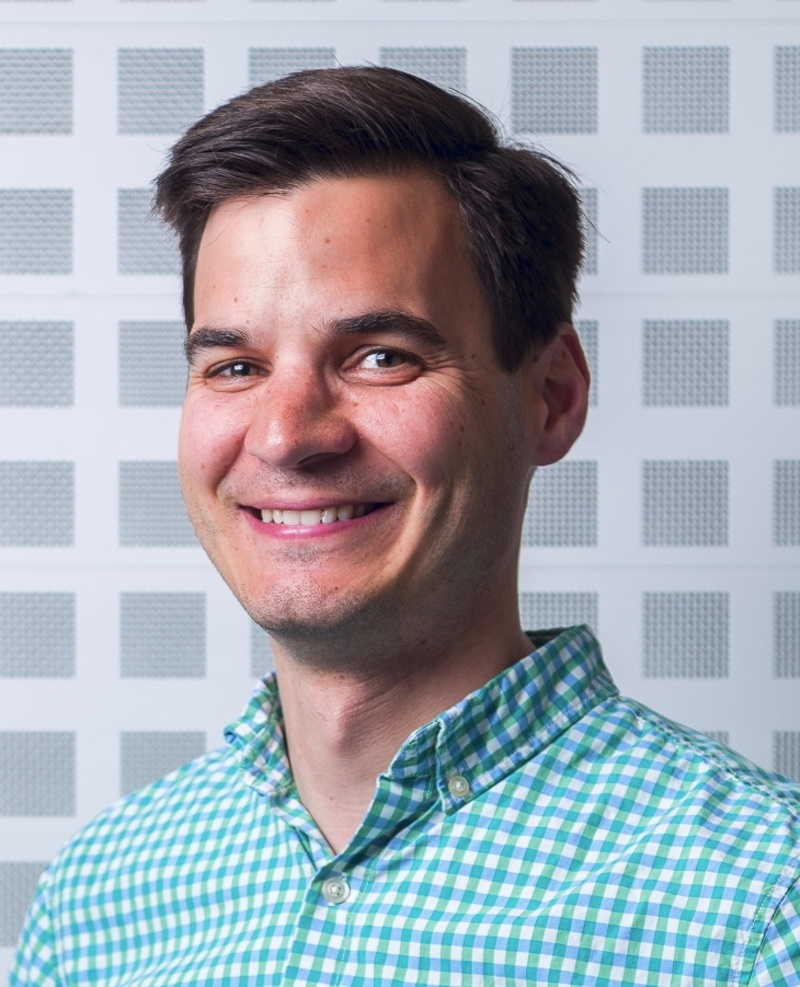}
\caption{{\textbf{Matthew Helgeson}~is currently Associate Professor of Chemical
Engineering at the University of California Santa Barbara, where he is a
faculty affiliate of the Material Research Laboratory and the BioPACIFIC
Materials Innovation Platform. He previously received a BS in Chemical
Engineering from Carnegie Mellon University (2004) and a PhD in Chemical
Engineering from the University of Delaware (2009), and performed
postdoctoral research at the Massachusetts Institute of Technology.
Prof. Helgeson's research focuses on rheological and structural
characterization of soft materials, including the development of methods
for probing fluid microstructure and dynamics out of equilibrium and
during processing.
{\label{333370}}%
}}
\end{center}
\end{figure}

\subsection*{GRAPHICAL ABSTRACT}\selectlanguage{english}
\begin{figure}[h!]
\begin{center}
\includegraphics[width=0.70\columnwidth]{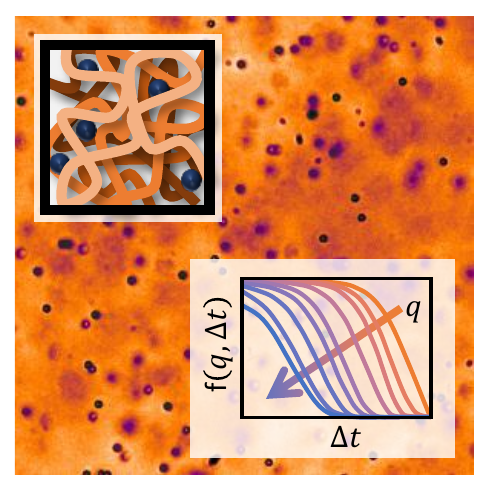}

\end{center}
\end{figure}

\selectlanguage{english}
\bibliography{converted_to_latex.bib}{}

\end{document}